# An inverse designed achromatic flat lens operating in the ultraviolet

SOURANGSU BANERJI,[1] AND BERARDI SENSALE-RODRIGUEZ[1,*]

[1]*Departmen of Electrical and Computer Engineering, The University of Utah, Salt Lake City, Utah, UT 84112, USA*
**berardi.sensale@utah.edu*

**Abstract:** We demonstrate an inverse designed achromatic, flat, polarization-insensitive diffractive optic element, i.e., multilevel diffractive lens (MDL), operating across a broadband range of UV light (250 nm – 400 nm) via numerical simulations. The simulated average on-axis focusing efficiency of the MDL is optimized to be as high as ~86%. We also investigate the off-axis focusing characteristics at different incident angles of the incoming UV radiation such that the MDL has a full field of view of 30˚. The simulated average off-axis focusing efficiency is ~67%, which is the highest reported till date for any chromatic or achromatic UV metalens or diffractive lens to the best of our knowledge. The designed MDL is composed of silicon nitride. The work reported herein will be useful for the miniaturization and integration of lightweight and compact UV optical systems.

## 1. Introduction

Optical devices operating in the ultraviolet regime of the electromagnetic spectrum have a myriad of applications in areas such as nanofabrication, military, medical, sterilization, etc. **[1, 2]**. From such perspective, it is important to develop efficient and cost-effective optical devices (and systems) to carry out these sensitive tasks with utmost precision and accuracy. Speaking of optical devices (or systems) in general, or even the ones operating in the UV, the most basic and principal element usually is a lens. For example, in the nanofabrication industry, right from shaping the incident light to facilitate writing masks, to reducing its spot size for etching out nanoscale features on these masks to resolving such features post-fabrication under a microscope, lenses form a significant part at every step of this process [**2**]. The same is also true for various military applications ranging from surveillance cameras to precision lenses for weaponry [**3**]. Finally, even for the medical industry, lenses are an integral part of any imaging or health monitoring system [**4**]. Therefore, it is important to appreciate the significance of lenses with the understanding that any significant progress made in the field of optical lens design which will help to mitigate the tasks that a lens does with higher efficiency, precision, accuracy, and portability at the expense of lower manufacturing and operational cost will be lucrative as well as revolutionary.

Traditional refractive lenses have been the go-to strategy to build these UV optical devices (or systems); and have performed reasonably well in terms of efficiency, precision, and accuracy; yet the idea of portability at a lower operating and manufacturing cost remains a potential challenge [**5**]. *Even though we firmly believe that these conventional refractive lenses will be the most practical way-to-go for a long time in the near future, it becomes necessary to look into alternatives, which, perhaps, could replace such lenses (or optical systems) in the long run one day or rather co-exist to facilitate a better approach in solving the existing problems.* With this philosophy, the idea or concept of flat optics evolved in the first place. Diffractive Optical Elements (DOEs) were the first of its kind, which, instead of harnessing the refractive phenomenon, relied on utilizing diffraction to guide incident light [**6**]. The ability to maintain a constant thickness at larger bending angles by decreasing the local period of the diffractive optic in contrast to a refractive one enabled these structures to be planar and lightweight. This potentially opened the field of diffractive optics, which now fulfilled all the above requirements in terms of efficiency, precision, accuracy, portability (due to being planar and lightweight) at a lower operational and manufacturing cost. From a technological standpoint, this was a considerable achievement, and the field witnessed a lot of progress [7, 8].

However, DOEs have also had their own set of problems associated with a drop in efficiency in high numerical aperture systems due to power being diverted to guided mode resonances instead of the propagating modes as well as their poor broadband performance due to significant chromatic aberrations [9, 10]. This prevented a large-scale industrial rollout for DOEs. Both problems are solved to some extent through parametric optimization of the geometry of constituent elements of the diffractive structure as well as through the utilization of harmonic phase shifts [11] and higher orders of diffraction [12]. However, such an approach is limited only to a discrete number of operating wavelengths. Recent demonstrations of DOEs based out of MDLs designed using the same principle of parametric optimization of its constituent elements have mitigated the problem of discrete operational bandwidth *with a modification of the phase transmittance function of a lens*. It displayed operation across a continuous bandwidth at both low and high numerical apertures with high efficiency [13, 14].

Design of achromatic multi-level diffractive lenses (MDLs) via computational optimization of the lens surface topography have now already been successfully demonstrated in the visible [13-19], near infrared [20], short-wave infrared, [21] long-wave infrared [22], THz [23, 24], and microwave [25] bands. Recently, we also highlighted the design of a single achromatic MDL operating across a continuous spectrum of wavelengths from 450nm to 15μm [26, 27] by utilizing the same "*phase as a free parameter*" concept as has been adopted in this paper. We have also successfully extended this concept to demonstrate other functionalities apart from achromaticity, where we have designed MDLs with a Field of View (FOV) up to 50° for wide-angle imaging and MDLs to render a Depth of Focus (DOF) imaging of up to 6 meters in the NIR [28]. Other noteworthy works include the design of MDLs to highlight broadband holograms enabling multi-plane image projection [29] as well as in holographic displays for AR and VR applications [30, 31]. In addition to this, the usefulness of MDLs in the construction of solar cells for efficient solar energy harvesting has also recently emerged [32-34]. MDLs have also found their way into the area of maskless lithography [35]. Finally, computational imaging with single and multi-aperture MDLs have also been demonstrated [36-38].

In this paper, we utilize the "*phase as a free parameter*" concept to demonstrate an inverse designed achromatic, flat, polarization-insensitive MDL operating across a broadband range of UV light (250 nm – 400 nm). The average on-axis and off-axis focusing efficiency of the MDL is ~86% and ~67%, respectively, with a FOV of 30°. The use of silicon nitride as the material of choice for the MDL design is what sets this study apart from any previous demonstration of MDLs. Moreover, recently in [27, 28], it was also showcased that an MDL can be designed to operate across an almost unlimited bandwidth; from that perspective, the work in this paper may just seem to be an extension of the bandwidth to a higher frequency range. Here, we seek to remind the reader that this is not entirely true. There are two reasons for this. First, the use of an MDL designed to operate across a vast bandwidth, as demonstrated in [28], would be an overkill for applications that only operate in the UV range. Second, upon close inspection in [28], one can observe that the efficiency falls on an average of up to ~70% in the lower wavelength range in contrast to an average of ~85% in the IR range. Therefore, it would be beneficial to choose MDLs, which are only designed to operate in the UV range with a higher focusing efficiency.

## 2. Theory and design

Our design methodology takes its inspiration from an old interdisciplinary field of information theory and optical images or "*optical information theory*" which dictates how imaging is inherently linked to information transfer from the object to the image plane back from the year 1955 [**39, 40**]. This can be accomplished in many ways ranging from lens-based to lens-less techniques [**28**]. The lens-based one-to-one mapping approach is preferred due to its high signal-to-noise ratio at each image point. In conventional imaging, when an incident wave impinges on a lens surface, it forms a focal spot at its focal plane. If we now assume that the main mode of incident field propagation from the lens plane to the focal plane is through diffraction instead of refraction, the field in the focal plane $U(x',y',\lambda,f)$ can be modeled with the Fresnel-Kirchhoff diffraction integral as:

$$U(x',y',\lambda,f) = \frac{e^{ikf}}{i\lambda f} \iint T(x,y,\lambda) \cdot e^{i\frac{k}{2f}[(x-x')^2+(y-y')^2]} dxdy, \quad (1)$$

where $f$ is the focal length, $k = \frac{2\pi}{\lambda}$, $(x,y)$ are coordinates in the lens plane, $(x',y')$ are coordinates in the focal plane, and $T(x,y,\lambda)$ is the pupil function of the lens. We assume that only the light that falls inside the MDLs active area is diffracted. Light that falls outside this area propagates though unaltered. The intensity in the focal plane is given by:

$$I(x',y',\lambda,f) = |U(x',y',\lambda,f)|^2. \quad (2)$$

We can express eqn. (1) as $U(x',y',\lambda,f) = P\{T(x,y,\lambda)\}$, where $P\{.\}$ is an operator that transforms the pupil function into the field in the focal plane. We note here that $P\{.\}$, in principle, is analytic, and the integral is bounded by the finite spatial extent of the pupil function. It is well known that $P\{.\}$ is invertible [**41**]. In other words, the field at the focal plane may be backpropagated to the lens plane. This is also evident from Maxwell's equations, which are time-reversible. Therefore, we can express the pupil function as

$$T(x,y,\lambda) = P^{-1}\{U(x',y',\lambda,f)\} = P^{-1}\{A(x',y',\lambda,f)e^{iB(x',y',\lambda,f)}\}, \quad (3)$$

where *A* and *B* are real-valued functions representing the amplitude and phase of the complex scalar field in the focal plane, respectively. While designing any suitable lens, the only restriction now is that $A = \text{sqrt}(I_{des})$, where $I_{des}$ is the desired intensity distribution of the focal spot. Hence, ***the phase of the field in the image or focal plane is a free parameter***. From (3), it is also clear that there will be one function representing *T* for each choice of *B*. Therefore, ***the pupil function of a lens, T, is not unique***.

A direct result of this can be gleaned from the case of a parabolic-phase pupil function, which converts an incident plane wave to a converging spherical wave. However, the most important observation here is that now one can add any arbitrary function riding on this parabola whose spatial frequencies are larger than those that will propagate in free space, without having any effect on the focal spot (if the focal length $\gg \lambda$). Therefore, one can simply modify the pupil function in infinitely different ways without modifying the focal spot; and essentially formulate it as an "*inverse design*" problem to solve for the appropriate pupil function *T* in eqn. (3) that satisfies the given design constraint, which in this paper is "*achromaticity*". This is a very common technique employed in earlier works by Lohmann [**42**], Wai-Hon Lee [**43**], and Lesem, Hirsch and Jordan [**44**] as well as extensively used in digital holography by Bryngdahl [**45**] and later on by others [**46**], where a hologram is designed to project a pre-defined intensity pattern and allow *the phase in the image plane to be arbitrary*.

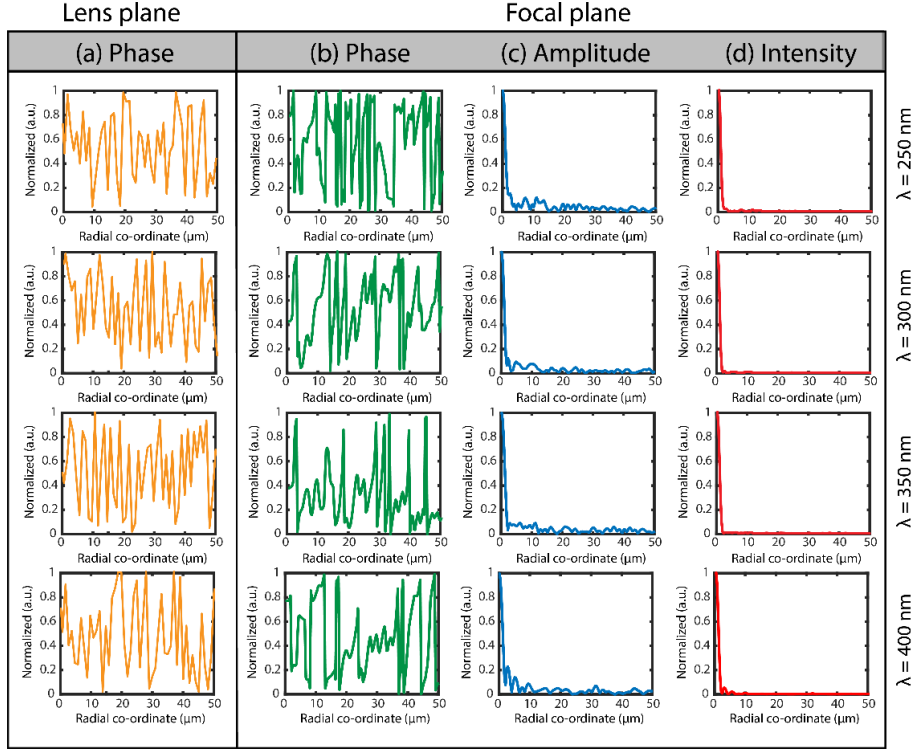

Fig. 1. Schematic of an incident field propagation from the lens plane to the focal plane of MDL is through diffraction (within the Fresnel regime). The MDL imparts its pupil function on the incident light. An image sensor discards the phase of the field in the focal plane, and only records the intensity. The pupil function varies with $\lambda$. The corresponding normalized (a) phase in the lens plane, and (b) phase, (c) amplitude, and (d) intensity distribution of the field in the focal plane with $\lambda = 250$ nm to $\lambda = 400$ nm (cross-section through the center) is shown.

This "*inverse design*" formulation can now be adapted for an MDLs phase transmittance function and solved with a computational optimization technique to solve for a (degenerate) lens-pupil function required to achieve achromatic focusing or any other desired MDL functionality. Fig. 1(a) illustrates this phase shift imposed on an incident plane wave by the MDL for $\lambda = 250$ nm to $\lambda = 400$ nm which is expressed as $\Psi = (2\pi/\lambda)*h*(n-1)$, where $h$ is the MDL ring height distribution and n is the refractive index at given $\lambda$. The MDL pupil function is then expressed as $e^{i\Psi}$ (which is analog to $T$ in eqn. (3) above). The phase and amplitude of the field distribution in the focal plane are plotted in Fig. 1(b-c) for $\lambda = 250$ nm to $\lambda = 400$ nm, respectively. All image sensors (or detectors) measure the square of the amplitude of the field distribution, i.e., the intensity distribution, and discard the phase, which is evidenced from the plot in Fig. 1(d). This shows that even though the phase distribution in the focal plane differs as a function of the wavelength, the intensity distributions are almost identical and simply scale with wavelength, resulting in a single-surface MDL that is achromatic from $\lambda = 250$ nm to $\lambda = 400$ nm.

### 3. Results and discussion

The computational optimization technique, which was chosen to determine the MDL's surface height profile, is a modified version of the direct binary search technique viz. Gradient Descent

Assisted Binary Search (GDABS) algorithm. Fig. 2(a) depicts the flow diagram of the algorithm. A complete description of the working principle of the GDABS algorithm is already provided in [23, 24]; therefore, we choose to omit an in-depth discussion of the same. However, to summarize it briefly, the algorithm starts with an initial random ring height profile for the DOE and moves ahead by traversing each ring at a time by either increasing or increasing its height by Δh. Next, the transmitted field, the diffracted field, and the Figure of Merit (FoM) are calculated. At the decision-making step, the algorithm evaluates a gradient of the FoM function to ensure a favorable path towards convergence. The FoM was chosen in a way to maximize the focusing efficiency of the MDL by selecting the distribution of heights of its constituent rings (see Fig. 2(b-c)). Mathematically, the FoM can be expressed as,

$$\text{FoM} = \eta_{on-axis} = \frac{\sum_{j=1}^{j=n} \frac{\int_{-\frac{3*w_{xj}}{2}}^{\frac{3w_{xj}}{2}} \int_{-\frac{3*w_{yj}}{2}}^{\frac{3w_{yj}}{2}} I_f(x',y',\lambda_j)}{\int_{-\frac{x}{2}}^{\frac{x}{2}} \int_{-\frac{y}{2}}^{\frac{y}{2}} I_i(x,y,\lambda_j)}}{m}, \qquad (4)$$

where $m$ equals the number of wavelength samples, variables $(w_{xj}, w_{yj})$ are the full width half maximum (FWHM) of the theoretical diffraction limited PSF in both the $x$- and $y$-direction for the respective wavelength sample. This FWHM value was calculated using the formula in [13]. In the numerator, $I_f(x',y',\lambda_j)$ is the intensity at the focal plane; whereas, in the denominator $I_i(x,y,\lambda_j)$ represents the intensity at the lens plane (rather total power impinging on the MDL). The fixed quantity $\left(\frac{3}{2}\right)$ in the term $\frac{3w_{xj}}{2}$ denotes that the optimization routine will try to maximize the intensity within a spot of diameter equal to 3 times the theoretical diffraction limited FWHM of the PSF at the focal plane. This is also the definition for focusing efficiency ($\eta_{on-axis}$). The optimization process terminates when the gradient became zero across an entire subsequent iteration, i.e., a local minimum is reached. Upon termination, the final output is the optimized ring height distribution for the MDL.

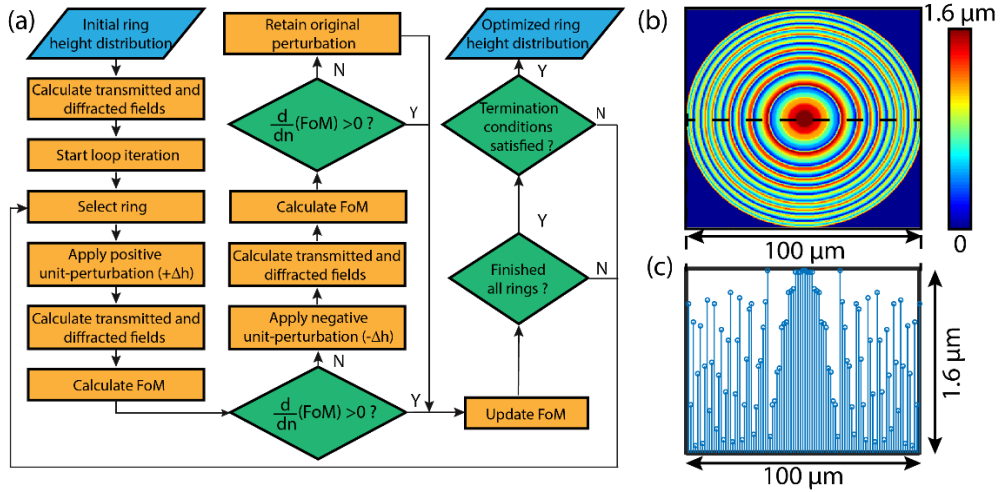

Fig. 2. (a) Flow Diagram of the Gradient Descent Assisted Binary Search (GDABS) Algorithm. This is the same algorithm as described in [23] but adapted to optimize for rotationally symmetric structures i.e. MDL ring heights. (b) Optimized ring height distribution of the designed MDL and its corresponding (c) cross-section across the middle of the structure. The maximum height ($h_{max}$) of the structure is 1.6μm.

The optimization routine was coupled with a conventional Fresnel–Kirchhoff diffraction integral to model the beam propagation (transmitted field and diffracted field at each point) starting from the lens plane up to the focal plane along the entire $z$ range. Specifically, the model assumes a non-paraxial unit amplitude uniform illumination to impinge on the MDL for both the on- and off-axis incidence. However, for oblique incidence, an additional phase term of the form:

$$e^{i\varphi} = e^{ik(xcos\theta + ysin\theta)}, \quad (5)$$

was added to the pupil function $T$ of eqn. (3) to account for the dependence of phase profiles.

The rotational symmetric nature of these MDL structures was utilized to speed up the computation time for the optimization routine. We acknowledge the fact here that advanced techniques like, for example, the adjoint method [**47, 48**], can also be realized to achieve results with similar computational complexity comparable to our GDABS technique. Nonetheless, our method lends itself to a simple and modular implementation of this "***inverse problem***" that enables the incorporation of multi-objective functions and fabrication constraints in a natural manner.

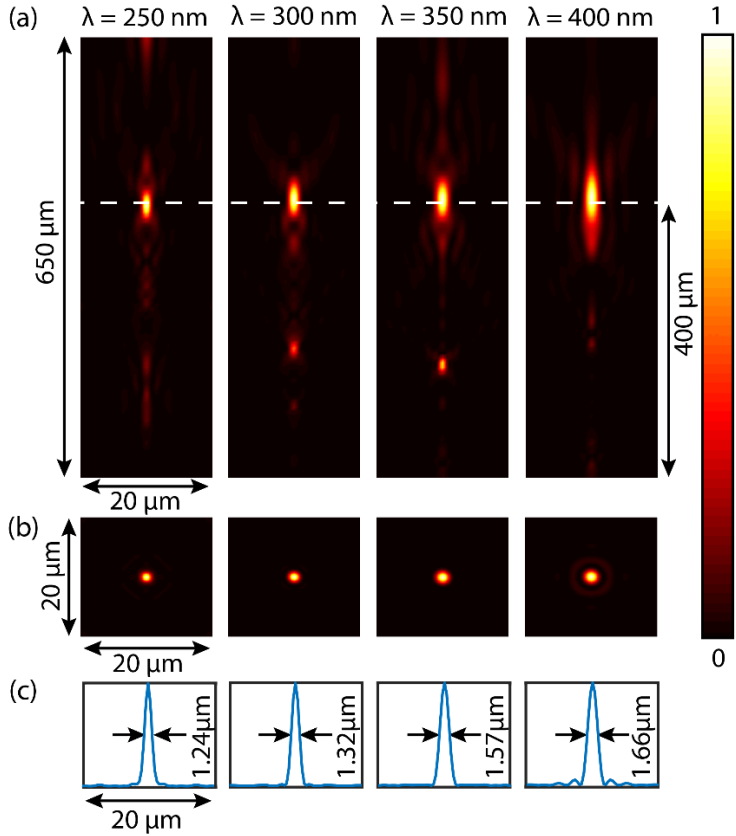

Fig. 3. (a) The $z$-propagation plot for $\lambda$ = 250 nm to $\lambda$ = 400 nm of the designed MDL at optimized wavelengths. Achromatic focusing capability is evidenced as each wavelength comes into focus at 400μm. The corresponding (b) 2D PSF plots and the (b) 1D cross-section of the PSFs across the middle.

The design parameters were chosen as follows: the rotationally symmetric MDL was designed to be 100μm in diameter, as shown in Fig. 2(b-c). As one can clearly see from the cross-section of the MDL in Fig. 2(c) that the surface profile of the MDL consists of multilevel rings having a maximum height ($h_{max}$) = 1.6 μm, and a unit ring height $\Delta h$ = 0.016 μm, which sets the number of distinct height levels (p) to p=100. Each constituent ring of the MDL has a width ($w$) = 1 μm. Therefore, the total number of constituent rings within each MDL = 50. The focal length of the designed MDL was fixed at 400 μm, therefore the numerical aperture (NA) = 0.1240 using the formula in [11]. Silicon nitride was the material of choice for this MDL design due to the material being transparent within the UV range. The refractive index values were extracted from [49-51]. The real part of the index of refraction was modeled through the following dispersion formula [50]:

$$n^2 - 1 = \frac{2.8939\lambda^2}{\lambda^2 - 0.13967^2}. \tag{6}$$

In the analyzed wavelength range absorption was neglected, as also done in previous works in the literature e.g. [52], since $h_{max}$ << penetration length in silicon nitride. The designed MDL can be fabricated starting from a silicon nitride film deposited on a silicon wafer through multi-step lithography and etching of the silicon nitride surface to define the MDL profile followed by removal of the handle wafer through DRIE and $XeF_2$ etch resulting ultimately on a lens consisting of a very thin patterned silicon nitride membrane [53, 54].

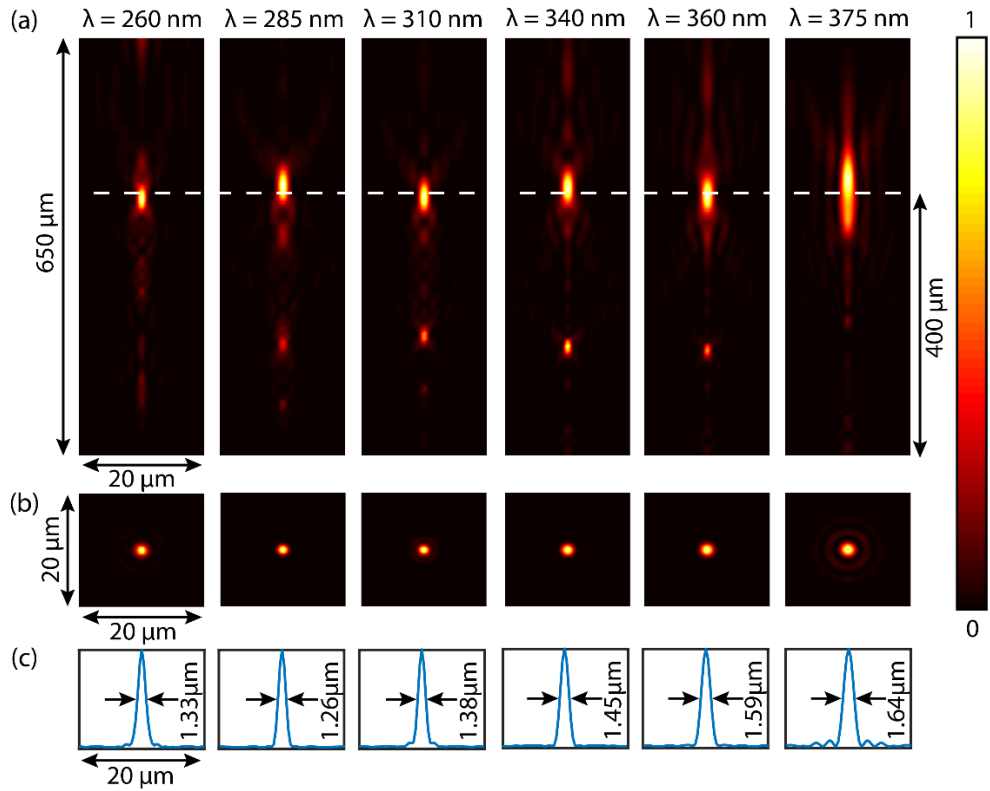

Fig. 4. (a) The z-propagation plot for $\lambda$ = 250 nm to $\lambda$ = 400 nm of the designed MDL at intermediate wavelengths i.e. wavelengths at which the MDL was not optimized for. The corresponding (b) 2D PSF plots and the (c) 1D cross-section of the PSFs across the middle for the same intermediate wavelengths.

The $z$-propagation plot, as shown in Fig. 3(a) for all the designed wavelengths portrays how the incident UV light traverses the optical path length in air before coming into focus at 400 µm. The simulated PSFs of Fig. 3(b) show achromatic focusing. Fig. 3(c) depicts the cross-section across the center of the simulated on-axis PSFs. We also investigated the designed MDL's performance at six representative intermediate wavelengths, i.e., 260nm, 285nm, 310nm, 340nm, 360nm and 375nm. Fig. 4(a-c) which shows the corresponding z-propagation, 2D and 1D cross-sections of the PSFs at these wavelengths. Achromaticity is still observed with negligible compromise in performance as evidenced from the 2D and 1D cross section of the PSF plots. To validate this, we have included a plot of the shift in focal length versus the incident wavelength in Fig. 5, which shows that the average chromatic shift is only ~1.2%. Upon close inspection of Fig. 5, we observe that the nominal shift is only at intermediate wavelengths for which the MDL has not been optimized. The average FWHM, as calculated from the PSFs, was 1.44 µm, which is close to the average theoretical diffraction limit of 1.30 µm.

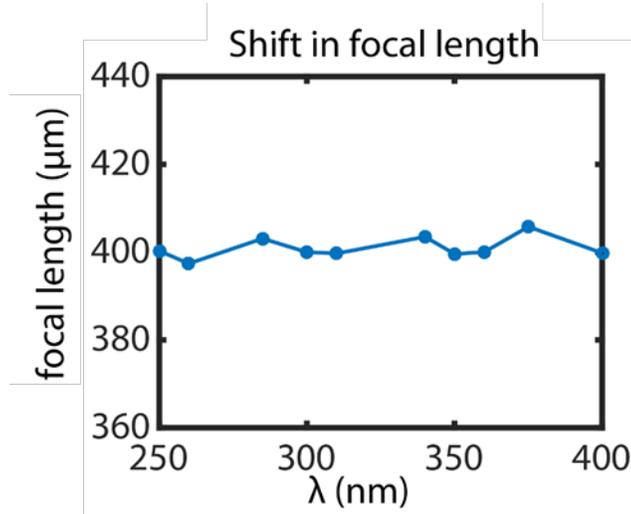

Fig. 5. The shift in focal length versus wavelength for the MDL with a designed focal length of $f$ = 400 µm. The average shift in focal length is ~1.2% across the entire broadband range.

We also simulated the off-axis focusing behavior of the designed MDL (Fig. 6) as a function of the half-angle ($\theta$) for the designed wavelengths. As can be observed from Fig. 3(b) and Fig. 6(a), there is a negligible difference in the 2D PSFs for the on and off-axis half-angle $\theta = 5º$. However, we do see a small amount of astigmatism at off-axis half-angle, $\theta = 10º$ for $\lambda$ = 300nm, and 350 nm. For off-axis half-angle, $\theta = 15º$, we see that the effect of astigmatism is more pronounced, which distorts the PSFs and would ultimately affect its imaging performance (distortion and blur). To verify this, we also carried out an aberration analysis based on Zernike polynomials in with the off-axis PSFs at $\theta = 15º$ under the broadband illumination. The fit was achieved with a least squares fit method. The indexing scheme used was Fringe. One can observe from the plot of Fig. 7 that astigmatism is indeed one of the dominant off-axis aberration for the designed MDL apart from piston and spherical aberrations. Coma and quadrafoil also contribute towards aberrations of the MDL. Therefore, we choose to limit our off-axis simulation for the MDL up to off-axis half-angle, $\theta = 15º$. This gives us a full FOV (2$\theta$) of 30º for the final design. The unit of angle for both the off-axis half angle ($\theta$) and off-axis full angle (2$\theta$) is in degrees (˚).

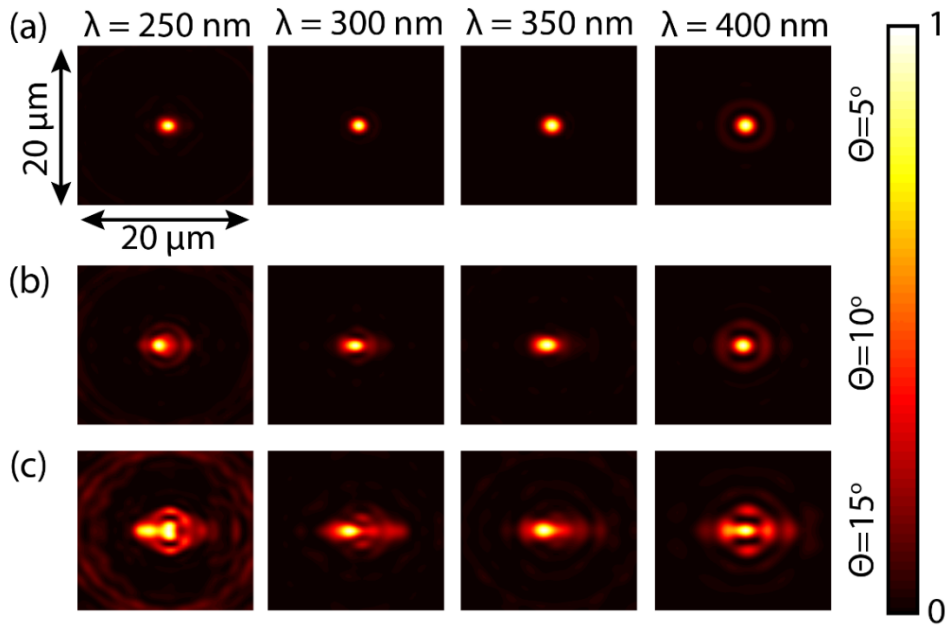

Fig. 6. The 2D PSF plots of the designed MDL for the half-angle off-axis cases of (a) θ = 5º, (b) θ = 10º and (c) θ = 15º for $\lambda$ = 250 nm to $\lambda$= 400 nm. Off-axis aberrations (predominantly astigmatism is observed from θ = 10º to θ = 15º. As a result, the imaging performance of the MDL would be, to an extent, compromised.

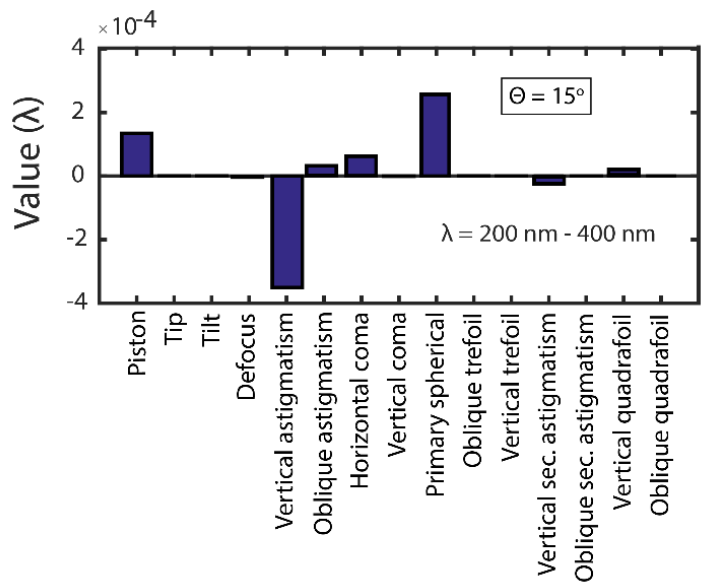

Fig. 7. Simulated coefficients of the constituent Zernike polynomials for the MDL at θ = 15º under the broadband illumination.

Next, we calculated the on- and off-axis focusing efficiency of the MDL. For the plot shown in Fig. 8(b), we performed a second averaging over the total number of half-angles (θ). The wavelength-averaged (250 nm to 400 nm) on-axis and off-axis focusing efficiency for the MDL is 85.53% and 67.13% respectively. Even for a low NA design as reported in this paper, we note that shadowing effects clearly affect focusing efficiencies for oblique incidence. This is evidenced from the PSF plots (appearance of additional fringes) of Fig. 6(b-c) and in Fig. 8(b) where the average efficiency decreases with increase in θ. From the plot of Fig. 8(b) we observe that the MDL operates with a loss of up to ~15% in efficiency across a full viewing angle (2θ) of up to 20⁰. Even though the simulated average on- and off-axis focusing efficiencies are predicted to be as high as 86% and 67%, respectively, one can expect this value to come down in experiments due to systemic and post fabrication errors in the MDL's ring heights and widths since the current optimization routine is not equipped to account for such errors during optimization [**23**]. Hence, it is quite important now to understand that incorporation of a suitable FoM metric (apart from the one currently used in this paper) in the optimization routine is important since it could ultimately help to design more robust, fabrication error tolerant and even wide FOV MDLs in the future for various other UV applications.

Table 1. Survey of broadband metalenses and diffractive lenses operating in the UV range

| Reference | Type | Material | Wavelength range | Numerical aperture | Average efficiency (Simulated) | Field of view (FOV) |
|---|---|---|---|---|---|---|
| **Guo et al. [55]** | metalens (chromatic) | AlN | 244 nm - 375 nm | > 0.1 | ~37% (on-axis) Not reported (off-axis) | Not reported |
| **Kanwal et al. [52]** | metalens (chromatic) | Si3N4 | 250 nm - 400 nm | 0.75 | ~55% (up to ~77% at one wavelength) (on-axis) Not reported (off-axis) | 5⁰ x 5⁰ |
| **Hu et al. [56]** | metalens (achromatic) | AlN | 234 nm - 274 nm | > 0.1 | ~44% (on-axis) Not reported (off-axis) | 10⁰ x 10⁰ |
| **Li et al. [57]** | diffractive lens | Sapphire | 360 nm – 370 nm | Not reported | ~81% (on-axis) Not reported (off-axis) | Not reported |
| **This work** | diffractive lens (achromatic) | Si3N4 | 250 nm - 400 nm | 0.124 | ~86% (on-axis) ~67% (off-axis) | 30⁰ x 30⁰ |

Speaking of focusing efficiency, we evaluated the performance of the designed MDL to some of the prominent metalens and diffractive lens designs in the scientific literature operating primarily in the UV range in Table 1. Regarding off-axis focusing efficiency, none of the previous works had reported any data. Moreover, we would like to point out here that not all the references given in Table 1 follow a consistent method for calculating focusing efficiency. In fact, most of the reference papers, do not even report how the authors defined focusing efficiency as in [**52, 55**]. In [**56**], the focusing efficiency was defined as "*Furthermore, focusing efficiency is a basic focusing performance metric, which is defined as the ratio of the light intensity in the area of FWHM in the focal spot and the total light intensity in the whole focal plane*". Finally, in [**57**], the focusing efficiency was defined as "*For N-level KPL, the efficiency is determined by,*

$$\eta(N) = \frac{sin^2\left(\frac{\pi}{N}\right)}{\left(\frac{\pi}{N}\right)^2}.$$

*For N = 2 and N = 4, theoretical efficiencies are 40.5% and 81.0%, respectively."*

From such perspective, we follow a conservative and uniform definition of focusing efficiency across published works [13, 20-22]. Moreover, from the definition of eqn. (3), one can easily comprehend the FWHM refers to a diffraction limited spot, meaning that diffraction efficiency is not improved by adding aberration. In terms of FOV, Kanwal et. al. [52] and Hu et. al. [56] reported a FOV of up to 5⁰ x 5⁰ and 10⁰ x 10⁰, respectively. Therefore, it is evident that the MDL design reported in this paper is amongst the highest in terms of both on- and off-axis focusing performance, FOV along with achromaticity across both metalenses and conventional diffractive lenses.

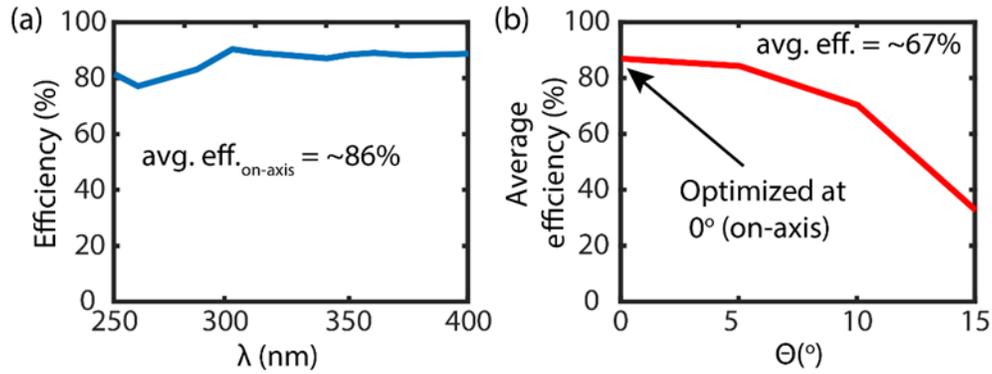

Fig. 8. (a) On-axis focusing efficiency of the MDL across a 150 nm bandwidth. (b) The average focusing efficiency across both on- and off-axis of the MDL as a function of the half-angle (θ) of the incident UV light.

## 4. Conclusion

In conclusion, we have successfully presented the design of an achromatic MDL operating across a broadband range of UV light (250 nm – 400 nm) via numerical simulation. We characterized and ascertained the simulated average on- and off-axis focusing efficiency of the MDL as ~86% and ~67%, respectively. The MDL has a FOV of 30˚. The designed MDL is composed of silicon nitride. The work reported herein will be useful in the miniaturization and integration of lightweight and compact UV optical systems.

**Funding**

National Science Foundation (NSF) (ECCS # 1936729, MRI #1828480).

**Disclosures**

The authors declare no conflicts of interest.